\RequirePackage{fix-cm}
\documentclass{svjour3}
\usepackage{amssymb}
\usepackage{amsmath}
\usepackage{graphicx}
\usepackage{subfigure}

\begin{document}
\title{Quantum entanglement for atoms coupling to fluctuating electromagnetic field in the cosmic string spacetime}
\author{Zhiming Huang}
\institute{
Zhiming Huang \at
School of Economics and Management, Wuyi University, Jiangmen 529020, China \\
\email{465609785@qq.com}\\
}
\maketitle
\begin{abstract}
We investigate entanglement dynamics for two atoms coupling with fluctuating electromagnetic field in the cosmic string spacetime. We calculate the entanglement for different conditions. It is found that the entanglement behaviors are dependent on vacuum fluctuation, spacetime topology, two-atom separation and atomic polarization orientation. After a long time of evolution, entanglement would vanish, which means entanglement affected by electromagnetic fluctuation can not maintain for a long time. For different spacetime topologies, entanglement presents different behaviors dependent on various parameters. When deficit angle parameter $\nu=1$ and atom-string distance is towards infinity, the results in flat spacetime are recovered. When atoms keep close to the string, entanglement can be improved; specially, when two atoms locate on the string and have no polarization of axial direction, atoms are not affected by the electromagnetic fluctuation and entanglement can remain unchanged. When two-atom separation is relatively large, entanglement exhibits oscillation behavior as atom-string distance varies. This indicates that the existence of string profoundly modifies on the vacuum fluctuation and atom-field interaction. In addition, when two-atom separation is small, entanglement gains better improvement. Many parameters and conditions provide us with greater freedom to control the entanglement behaviors. In principle, this is useful to sense the cosmic string spacetime topology structure and property, and discriminate different kinds of spacetime.
\end{abstract}
\section{Introduction}
Quantum entanglement is most fundamental difference between quantum classical world and quantum world, which has many amazing potential applications, such as quantum information processing and quantum computation \cite{Nilsen2000,Horodecki2009}. However, the ineluctable interaction between quantum system and environment may damage quantum entanglement, such as atom-field interaction. Recently, the entanglement dynamics is investigated for atoms coupling to fluctuating scalar field or electromagnetic field in Minkowski spacetime, and the results show that entanglement presents behaviors of degradation, generation, revival and enhancement under the influence of vacuum fluctuation \cite{Hu2015,Yang2016}. Lately entanglement dynamics for atoms affected by quantum field in curved spacetime are studied \cite{Hu2013,Huang2017a}, and the results shows that entanglement behaviors can be used to distinguish different spacetime backgrounds.

On the other hand, the cosmic string spacetime with nontrivial topological defect is a typical curved spacetime. The topological defect of cosmic string is originated from the evolution of the early universe due to the symmetry breaking phase transitions \cite{Copeland2011,Hindmarsh2011}. The cosmic string formed under certain conditions can be stable and can exist for a long time until now, which may provide us with information about particle physics and the early universe. However, due to its inherent complexity, it is not easy to understand the nature of the cosmic string. Cosmic string have energy momentum tensor and it can induce some interesting gravitational effects, such as gravitational lensing effect \cite{Vilenkin1984,Gott1985} and gravitational Aharonov-Bohm effect \cite{Ford1981,Bezerra1989}. For better understanding the nature of the cosmic string, some quantum processes have been investigated in the cosmic string spacetime, such as radiative properties \cite{Cai2015,Zhou2016,Cai2018}, geometric phase \cite{Bakke2008,Cai2018a} and Casimir effect \cite{Bezerra2012a}.

Recently, we have examined the entanglement dynamics for two accelerating atoms interacting with a massless scalar field in the cosmic string spacetime \cite{Huang2020}. In comparison with massless scalar field, the fluctuating vacuum electromagnetic field is a more realistic model. In this work, we intend to analyze the entanglement behaviors for two identical atoms immersing in fluctuating electromagnetic field in the background of the cosmic string spacetime. We are intriguing in how the nontrivial spacetime topology affects the atom-field interaction and the entanglement behaviors, and whether entanglement behaviors can be exploited to detect the spacetime character and differentiate various spacetime. Our exploration would offer new insight for the property of the topological defect universe and the quantum field theory of curved spacetime.

This paper is organized as follows. In Section \ref{S2}, we firstly introduce interaction model between two two-level atoms and a quantum electromagnetic field, and the entanglement measure. And then in Section \ref{S3}, we do relevant calculation of entanglement quantity and discuss the entanglement behaviors in different circumstances. Finally, we present a summary of this paper in Section \ref{S4}. The natural units $\hbar = c = 1$ is adopted in this paper.
\section{Preliminaries}\label{S2}
Let us consider two identical two-level atoms weakly coupling with fluctuating electromagnetic field.
The Hamiltonian of the whole system can be written as
\begin{align}
H=H_{S}+H_{F}+H_{I}.
\end{align}
 $H_{S}$ is the Hamiltonian of two atoms, which takes the form
 \begin{align}
H_S=H_S^{(1)}+H_S^{(2)}, \quad H_S^{(\alpha)}=\frac{\omega}{2}\sigma_{3}^{(\alpha)}, \quad(\alpha=1,2),
\end{align}
where $\sigma_{i}^{(1)}=\sigma_{i}\otimes\sigma_{0}$,  $\sigma_{i}^{(2)}=\sigma_{0}\otimes\sigma_{i}$, with $\sigma_{i}$ $(i=1,2,3)$ being the Pauli operators and $\sigma_{0}$ being the $2\times2$ identity operator, and $\omega$ denotes the energy level spacing. $H_{F}$ is the Hamiltonian of the electromagnetic field. $H_{I}$ is the dipole interaction Hamiltonian between atoms and electromagnetic field, which has the form
 \begin{align}
H_I(\tau)=-\textbf{D}^{(1)}(\tau)\cdot\textbf{E}(x^{(1)}(\tau))
          -\textbf{D}^{(2)}(\tau)\cdot\textbf{E}(x^{(2)}(\tau)),
\end{align}
where $\tau$ is the proper time of the atoms, $\textbf{D}^{(\alpha)}(\tau)~(\alpha=1,2)$ is the electric-dipole moment of the atom, and ${\bf E}(x^{(\alpha)}(\tau))$ denotes the electric-field.
In the interaction representation the atomic dipole operators take the form $\textbf{D}^{(\alpha)}(\tau)=\textbf{d}^{(\alpha)}\sigma_{-} e^{-\rm{i}{\omega}{\tau}}+\textbf{d}^{(\alpha)*}\sigma_{+} e^{\rm{i}{\omega}{\tau}}$
with $\textbf{d}^{(\alpha)}=\langle 0|\textbf{D}^{(\alpha)}|1\rangle$.

We assume initially there is no correlation between the atoms and the electromagnetic field. Then the initial state of the total system can be written as $\rho_{tot}(0)=\rho(0)\otimes|0\rangle\langle0|$, where $\rho(0)$ is the initial state of the atomic system and $|0\rangle$ is the vacuum state of the electromagnetic field. In the weakly coupling limit, the evolution process of the atoms can be described by the Kossakowski-Lindblad master equation \cite{Gorini1976,Lindblad1976,Breuer2002}
\begin{align}
\frac{\partial\rho(\tau)}{\partial \tau}=-\rm{i}[H_{\text{eff}},\rho(\tau)]+\mathcal{L}[\rho(\tau)],\label{m1}
\end{align}
where
\begin{align}
H_{\text{eff}}=H_{S}-\frac{\rm{i}}{2}\sum_{\alpha,\beta=1}^{2}\sum_{i,j=1}^{3}H_{ij}^{(\alpha\beta)}\sigma_{i}^{(\alpha)}\sigma_{j}^{(\beta)},
\end{align}
and
\begin{align}
\mathcal{L}[\rho]=\frac{1}{2}\sum_{\alpha,\beta=1}^{2}\sum_{i,j=1}^{3}S^{(\alpha\beta)}_{ij}[2\sigma_{j}^{(\beta)}\rho\sigma_{i}^{(\alpha)}-\sigma_{i}^{(\alpha)}\sigma_{j}^{(\beta)}\rho-\rho\sigma_{i}^{(\alpha)}\sigma_{j}^{(\beta)}].\label{m2}
\end{align}
$S_{ij}^{(\alpha\beta)}$ and $H_{ij}^{(\alpha\beta)}$ are determined by the electromagnetic field correlation functions:
\begin{align}
G^{(\alpha\beta)}_{kl}(\tau-\tau')
=\langle0| E_{k}(\tau,\mathbf{x}_{\alpha}) E_{l}(\tau',\mathbf{x}_\beta) |0\rangle.
\end{align}
The Fourier transforms and Hilbert transforms of field correlation function are defined as
\begin{align}
&\mathcal{G}^{(\alpha\beta)}_{kl}(\lambda)=\int_{-\infty}^{\infty}d\Delta \tau e^{\rm{i} \lambda \Delta \tau}G^{(\alpha\beta)}_{kl}(\Delta \tau),\label{ft}\\
&\mathcal{K}^{(\alpha\beta)}_{kl}(\lambda)=\frac{P}{\pi \rm{i}}\int_{-\infty}^{\infty}d\varpi\frac{\mathcal{G}^{(\alpha\beta)}_{kl}(\varpi)}{\varpi-\lambda},
\end{align}
where $P$ is the principal value. The Kossakowski matrix $S_{ij}^{(\alpha\beta)}$ can be expressed analytically as
\begin{align}
S_{ij}^{(\alpha\beta)}
= A^{(\alpha\beta)}\delta_{ij}
 -iB^{(\alpha\beta)}\epsilon_{ijk}\,\delta_{3k}
 -A^{(\alpha\beta)}\delta_{3i}\,\delta_{3j} ,\label{CC}
\end{align}
where
\begin{align}
A^{(\alpha\beta)}
={\frac{1}{4}}\,[\,{\cal G}^{(\alpha\beta)}(\omega)
 +{\cal G}^{(\alpha\beta)}(-\omega)] ,\nonumber\\
B^{(\alpha\beta)}
={\frac{1}{4}}\,[\,{\cal G}^{(\alpha\beta)}(\omega)
 -{\cal G}^{(\alpha\beta)}(-\omega)] ,\label{AB}
\end{align}
with
\begin{align}
{\cal G}^{(\alpha\beta)}(\omega)
=\sum_{k,l=1}^3\, d^{(\alpha)}_k d^{(\beta)*}_l\,
{\cal G}^{(\alpha\beta)}_{kl}(\omega) .\label{GC}
\end{align}
Similarly, $H^{(\alpha\beta)}_{ij}$ can be obtained by replacing ${\cal G}^{(\alpha\beta)}_{ij}$ with ${\cal K}^{(\alpha\beta)}_{ij}$ in the above expressions.

We employ Wootters concurrence \cite{Wootters1998} as the entanglement measure. For two-qubit X-shape states
\begin{align}
\rho_{X}=\frac{1}{4}(I\otimes I+\sum_{i}^{3}c_{i}\sigma_{i}\otimes \sigma_{i}+c_{4}I\otimes\sigma_{3}+c_{5}\sigma_{3}\otimes I),\label{xstate}
\end{align}
with $-1\leq c_{i}\leq 1$,
the concurrence is given by
\begin{align}
E = \max(0,F_{1},F_{2}), \label{CCR}
\end{align}
where $F_{1} =\frac{1}{2} [| c_1+c_2| -\sqrt{(c_3-c_4-c_5+1) (c_3+c_4+c_5+1)}]$ and $F_{2} = \frac{1}{2} [| c_1-c_2| -\sqrt{(c_3+c_4-c_5-1) (c_3-c_4+c_5-1)}]$.
\section{Dynamics of quantum entanglement in the cosmic string spacetime}\label{S3}
In this section, we examine the entanglement dynamics for two-atom system coupled to electromagnetic vacuum fluctuation in the cosmic string spacetime.
Let us assume a straight and infinitely long comics string lies at the $z$ axis, the spacetime metric in the cylindrical coordinate $(t,r,\theta,z)$ is described as
 \begin{align}
ds^2=dt^2-dr^2-r^2d\theta^2-dz^2,\label{csm}
\end{align}
which looks like Minkowski spacetime metric, except that the azimuth angle $\theta$ is restricted in the range $0\leq\theta<{2\pi/\nu}$, $\nu=(1-4G \mu)^{-1}$ with $G$ and $\mu$ denoting the Newton gravitational constant and the string\rq s linear mass density respectively. The metric is a simple solution of the Einstein equations. The metric describes a locally flat but not globally flat spacetime with a planar angle deficit $\delta\theta=8\pi G \mu$.

By some calculations and derivations,  the operator of electromagnetic field in the cosmic string spacetime can be written as \cite{Cai2015}
 \begin{align}
A_{\xi}(x)=\int d\mu_j[c_{\xi j}(t) f_{\xi j}(x)+\text{H.c.}],\label{cs-field-operator}
\end{align}
where
 \begin{align}
\int d\mu_j=\sum^{\infty}_{n=-\infty}\int^{\infty}_{-\infty}d\kappa\int^{\infty}_0dk_{\perp}k_{\perp},
\end{align}
with $j\equiv\{n,\kappa,k_{\perp}\}$,
$c_{\xi j}(t)=c_{\xi j}(0)e^{-i\omega_j t}$ are the annihilation operator  of the field, and
 \begin{align}
f_{\xi j}(x)=\frac{1}{2\pi}\sqrt{\frac{\nu}{2\omega_j}}e^{i\kappa z}e^{i\nu n\theta}J_{|\nu n+\xi|}(k_{\perp}r)\;,\label{csmode}
\end{align}
with $\xi\in\{0,1,-1,3\}$, $\kappa\in(-\infty,\infty)$, $n\in Z$, $k_{\perp}\in[0,\infty)$, $\omega_j=\sqrt{\kappa^2+k^2_{\perp}}$ and $J_{|\nu n+\xi|}(k_{\perp}r)$ being the Bessel function.
 After some calculations, the correlation function can be expressed as \cite{Cai2015}
  \begin{align}
\langle0|E_k(x)E_l(x')|0 \rangle=\partial_k\partial_l^\prime\langle0|A_0(x)A_0(x')|0\rangle+\partial
_0\partial_0^\prime\langle0|A_k(x)A_l(x')|0\rangle,\label{green}
  \end{align}
 where
 \begin{align}
&A_r(x)=\frac{1}{\sqrt{2}}\int d\mu_j[(c_{+ j}(t) f_{+j}(x)+c_{-j}(t) f_{-j}(x))+\text{H.c.}],\\
&A_\theta(x)=-\frac{i r}{\sqrt{2}}\int d\mu_j[(c_{+j}(t)f_{+ j}(x)-c_{-j}(t) f_{-j}(x))-\text{H.c.}],\\
&A_{z,t}(x)=\int d\mu_j[c_{3j,0j}(t)f_{0 j}(x)+\text{H.c.}].
  \end{align}
The trajectories of two static atoms parallel to string can be expressed as
\begin{align}
t_1(\tau)&=\tau, &
\theta_1(\tau)&=0 , &
r_1(\tau)&=r , &
z_1(\tau)&=0 ,\nonumber\\
t_2(\tau)&=\tau, &
\theta_2(\tau)&=0 , &
r_2(\tau)&=r , &
z_2(\tau)&=L .\label{tt}
\end{align}
Substituting the atomic trajectories into field correlation function (\ref{green}), for $\alpha=\beta$, we obtain
\begin{align}
G_{11}^{(11)}(\tau-\tau')=&G_{11}^{(22)}(\tau-\tau')=\frac{\nu}{8\pi^{2}}\int d\mu_j e^{-i\omega_j \Delta \tau}\nonumber\\
&\times[\frac{\omega_j}{2}(J^{2}_{|\nu n+1|}(k_{\perp}r)+J^{2}_{|\nu n-1|}(k_{\perp}r))-\frac{1}{\omega_j}(\frac{\partial J_{|\nu n|}(k_{\perp}r)}{\partial r})^{2}],\nonumber\\
G_{22}^{(11)}(\tau-\tau')=&G_{22}^{(22)}(\tau-\tau')=\frac{\nu r^{2}}{8\pi^{2}}\int d\mu_j e^{-i\omega_j \Delta \tau},\nonumber\\
&\times[\frac{\omega_j}{2}(J^{2}_{|\nu n+1|}(k_{\perp}r)+J^{2}_{|\nu n-1|}(k_{\perp}r))-\frac{1}{\omega_j}\frac{\nu^{2}n^{2}}{r^{2}}J^{2}_{|\nu n|}(k_{\perp}r)],\nonumber\\
G_{33}^{(11)}(\tau-\tau')=&G_{33}^{(22)}(\tau-\tau')=\frac{\nu}{8\pi^{2}}\int d\mu_j e^{-i\omega_j \Delta \tau}\frac{k_{\perp}^{2}}{\omega_j}J^{2}_{|\nu n|}(k_{\perp}r),
\end{align}
where $\Delta \tau=\tau-\tau'$. For $\alpha\neq\beta$, we obtain
\begin{align}
G_{11}^{(12)}(\tau-\tau')=&G_{11}^{(21)}(\tau-\tau')=\frac{\nu}{8\pi^{2}}\int d\mu_j \cos(\kappa L)e^{-i\omega_j \Delta \tau}\nonumber\\
&\times[\frac{\omega_j}{2}(J^{2}_{|\nu n+1|}(k_{\perp}r)+J^{2}_{|\nu n-1|}(k_{\perp}r))-\frac{1}{\omega_j}(\frac{\partial J_{|\nu n|}(k_{\perp}r)}{\partial r})^{2}],\nonumber\\
G_{22}^{(12)}(\tau-\tau')=&G_{22}^{(21)}(\tau-\tau')=\frac{\nu r^{2}}{8\pi^{2}}\int d\mu_j\cos(\kappa L) e^{-i\omega_j \Delta \tau}\nonumber\\
&\times[\frac{\omega_j}{2}(J^{2}_{|\nu n+1|}(k_{\perp}r)+J^{2}_{|\nu n-1|}(k_{\perp}r))-\frac{1}{\omega_j}\frac{\nu^{2}n^{2}}{r^{2}}J^{2}_{|\nu n|}(k_{\perp}r)],\nonumber\\
G_{33}^{(12)}(\tau-\tau')=&G_{33}^{(21)}(\tau-\tau')=\frac{\nu}{8\pi^{2}}\int d\mu_j \cos(\kappa L)e^{-i\omega_j \Delta \tau}\frac{k_{\perp}^{2}}{\omega_j}J^{2}_{|\nu n|}(k_{\perp}r),\nonumber\\
G_{13}^{(12)}(\tau-\tau')=&-G_{13}^{(21)}(\tau-\tau')=-G_{31}^{(12)}(\tau-\tau')=G_{31}^{(21)}(\tau-\tau')=\nonumber\\
&\frac{\nu}{8\pi^{2}}\int d\mu_j \sin(\kappa L)e^{-i\omega_j \Delta \tau}\frac{\kappa}{\omega_j}J_{|\nu n|}(k_{\perp}r)\frac{\partial J_{|\nu n|}(k_{\perp}r)}{\partial r}.
\end{align}
The rest of terms are zero.
According to the residue theorem, the Fourier transforms (\ref{ft}) of the two point functions can be obtained. For $\alpha=\beta$, we obtain
\begin{align}
\mathcal{G}^{(11)}_{ij}(\omega)=\mathcal{G}^{(22)}_{ij}(\omega)=\frac{\omega ^3 }{3 \pi }f^{(11)}_{ij},
\end{align}
where
\begin{align}
&f^{(11)}_{11}=\frac{3\nu}{4}\sum_{n=-\infty}^{\infty}\int_0^1d\eta\frac{\eta  [\eta ^2 J_{| \nu  n| -1}(\omega  r \eta ) J_{| \nu  n| +1}(\omega  r \eta )+(2-\eta ^2) J^2_{| n \nu +1| }(\omega  r \eta ){}]}{ \sqrt{1-\eta ^2}}  ,\nonumber\\
&f^{(11)}_{22}=\frac{3\nu}{4}\sum_{n=-\infty}^{\infty}\int_0^1d\eta\frac{\eta  [-\eta ^2 J_{| \nu  n| -1}(\omega  r \eta ) J_{| \nu  n| +1}(\omega  r \eta )+(2-\eta ^2) J^2_{| n \nu +1| }(\omega  r \eta ){}]}{ \sqrt{1-\eta ^2}} ,\nonumber\\
&f^{(11)}_{33}=\frac{3\nu}{2}\sum_{n=-\infty}^{\infty}\int_0^1d\eta\frac{\eta ^3 J^2_{| \nu  n| }(\omega  r \eta ){}}{\sqrt{1-\eta ^2}},
\end{align}
and the rest of components are zero.
For $\alpha \neq \beta$,
\begin{align}
\mathcal{G}^{(\alpha\beta)}_{ij}(\omega)=\frac{\omega ^3 }{3 \pi }g^{(\alpha\beta)}_{ij},
\end{align}
where
\begin{align}
g^{(12)}_{11}=g^{(21)}_{11}=&\frac{3\nu}{4}\sum_{n=-\infty}^{\infty}\int_0^1 d\eta \frac{\cos(\omega L\sqrt{1-\eta ^2})\eta  }{ \sqrt{1-\eta ^2}} \nonumber\\
&\times[\eta ^2 J_{| \nu  n| -1}(\omega  r \eta ) J_{| \nu  n| +1}(\omega  r \eta )+(2-\eta ^2) J^2_{| n \nu +1| }(\omega  r \eta ){}],\nonumber\\
g^{(12)}_{22}=g^{(21)}_{22}=&\frac{3\nu}{4}\sum_{n=-\infty}^{\infty}\int_0^1 d\eta\frac{\cos(\omega L\sqrt{1-\eta ^2}) \eta }{ \sqrt{1-\eta ^2}} \nonumber\\
&\times[-\eta ^2 J_{| \nu  n| -1}(\omega  r \eta ) J_{| \nu  n| +1}(\omega  r \eta )+(2-\eta ^2) J^2_{| n \nu +1| }(\omega  r \eta ){}],\nonumber\\
g^{(12)}_{33}=g^{(21)}_{33}=&\frac{3\nu}{2}\sum_{n=-\infty}^{\infty}\int_0^1 d\eta\frac{\cos(\omega L\sqrt{1-\eta ^2})\eta ^3 J^2_{| \nu  n| }(\omega  r \eta ){}}{\sqrt{1-\eta ^2}} ,\nonumber\\
g^{(12)}_{13}=g^{(21)}_{31}=&-g^{(12)}_{31}=-g^{(21)}_{13}=\frac{3\nu}{4}\sum_{n=-\infty}^{\infty}\int_0^1 d\eta{\sin(\omega L\sqrt{1-\eta ^2})\eta ^2 J_{|\nu n| }(\omega  r \eta )}\nonumber\\
&\times[J_{| \nu  n| -1}(\omega  r \eta ) -J_{| \nu  n| +1}(\omega  r \eta )],
\end{align}
and the other terms are zero. For obtaining the above results, we have utilized the following properties of Bessel function \cite{Andrews1999}
\begin{align}
&\sum_{n=-\infty}^{\infty}J^{2}_{|\nu n+1|}(x)= \sum_{n=-\infty}^{\infty}J^{2}_{|\nu n-1|}(x) ,\nonumber\\
&\sum_{n=-\infty}^{\infty}J^{2}_{|\nu n|+1}(x)+ \sum_{n=-\infty}^{\infty}J^{2}_{|\nu n|-1}(x)=2\sum_{n=-\infty}^{\infty}J^{2}_{|\nu n+1|}(x), \quad (\nu\geq1) \nonumber\\
& J_{n-1}(x)+ J_{n+1}(x)=\frac{2n}{x}J_{n}(x),\quad J_{n-1}(x)- J_{n+1}(x)=2\frac{\text{d}J_{n}(x)}{dx}.
\end{align}

According to Eqs. (\ref{AB}) and (\ref{GC}), we can rewrite Eq. (\ref{CC}) as
\begin{align}
&S_{ij}^{(11)}=A\,\delta_{ij}-iA\epsilon_{ijk}\,\delta_{3k} -A\delta_{3i}\,\delta_{3j} ,\nonumber\\
&S_{ij}^{(22)}=B\,\delta_{ij}-iB\epsilon_{ijk}\,\delta_{3k} -B\delta_{3i}\,\delta_{3j} ,\nonumber\\
&S_{ij}^{(12)}=S_{ij}^{(21)}=C\,\delta_{ij}-iC\epsilon_{ijk}\,\delta_{3k} -C\delta_{3i}\,\delta_{3j},
\end{align}
with
\begin{align}
&A=\frac{\Gamma}{4} \sum_{i=1}^3 f^{(11)}_{ii}|\hat{d}^{(1)}_i|^{2} ,\nonumber\\
&B=\frac{\Gamma}{4} \sum_{i=1}^3 f^{(11)}_{ii}|\hat{d}^{(2)}_i|^{2} ,\nonumber\\
&C=\frac{\Gamma}{4} \sum_{i}^3 \sum_{j}^3g^{(12)}_{ij}\hat{d}^{(1)}_i\hat{d}^{(2)\ast}_i, \label{fts1}
\end{align}
where $\Gamma=  \omega^3 |\mathbf{d}|^2 /3\pi$ is the atomic spontaneous emission rate, and $\hat{d}^{(\alpha)}_{i}=d^{(\alpha)}_{i}/|\mathbf{d}|$ is a unit vector with $d^{(\alpha)}_{i}$ being the electric dipoles of the atoms. We assume that $|\mathbf{d}^{(1)}|=|\mathbf{d}^{(2)}|=|\mathbf{d}|$.

Now, let us derive the master equation (\ref{m1}) in the interaction representation. Any two-qubit states can be represented with Pauli operators
\begin{align}
\rho =\frac{1}{4} \sum _{i=0}^3 \sum _{j=0}^3 p_{i,j}(\tau) \sigma_{i}\otimes\sigma_{j}. \label{tes}
\end{align}
Here we suppose two atoms initially are prepared in X-shape state, combining Eq. (\ref{tes}) with Eq. (\ref{m1}), we can obtain the non-trivial coupled differential equations
\begin{align}
&p_{0,0}'(\tau)=0,\label{de1}\\
&p_{1,1}'(\tau)=2 [-(A+B) p_{1,1}(\tau )+C p_{0,3}(\tau )+C \left(p_{3,0}(\tau )+2 p_{3,3}(\tau )\right)],\nonumber\\
&p_{0,3}'(\tau)=-2 [2 B p_{0,0}(\tau )+2 B p_{0,3}(\tau )+C \left(p_{1,1}(\tau )-p_{2,2}(\tau )\right)],\nonumber\\
&p_{2,2}'(\tau)=-2 (A+B) p_{2,2}(\tau )-2 C \left(p_{0,3}(\tau )+p_{3,0}(\tau )\right)-4 C p_{3,3}(\tau ),\nonumber\\
&p_{3,0}'(\tau)=-2 \left(2 A p_{0,0}(\tau )+2 A p_{3,0}(\tau )+C p_{1,1}(\tau )-C p_{2,2}(\tau )\right),\nonumber\\
&p_{3,3}'(\tau)=-4 \left(A p_{0,3}(\tau )+B p_{3,0}(\tau )\right)-4 (A+B) p_{3,3}(\tau )+4 C (p_{1,1}(\tau )- p_{2,2}(\tau )).\label{de2}
\end{align}
From above equations, we know that the X-shape evolution state is obtained. With Eq. (\ref{CCR}), we can get concurrence
\begin{align}
E=\max [0,\frac{1}{2}| p_{1,1}(\tau)+p_{2,2}(\tau)| -T_1,\frac{1}{2} | p_{1,1}(\tau)-p_{2,2}(\tau)| -T_2],\label{N1}
\end{align}
with  $T_1=\frac{1}{2} \sqrt{(1+p_{0,3}(\tau)-p_{3,0}(\tau)-p_{3,3}(\tau)) (1-p_{0,3}(\tau)+p_{3,0}(\tau)-p_{3,3}(\tau))}$ and $T_2=\frac{1}{2} \sqrt{(1-p_{0,3}(\tau)-p_{3,0}(\tau)+p_{3,3}(\tau)) (1+p_{0,3}(\tau)p_{3,0}(\tau)+p_{3,3}(\tau))}$.

Firstly let us examine the case of equilibrium state that can be obtained by setting change rate of Eq. (\ref{de2}) to be zero provided that two atoms are not very close to each other and not very near the string. The
linear equations can be solved and get $p_{1,1}(\infty )= p_{2,2}(\infty )= 0,p_{0,3}(\infty )=p_{3,0}(\infty )= -1,p_{3,3}(\infty )= 1$. From these solutions, we know that the state after a long time evolution is a separable state $|11\rangle$, which is independent of two-atom distance, atom-string distance, atom polarization and deficit angle parameter. This implies that two atoms can not get entangled for a long evolution time whatever the initial state of two atom is. In the following discussion, we choose the Werner state $p|\phi\rangle\langle\phi|+(1-p)\frac{I}{4}$ ($|\phi\rangle=\frac{1}{\sqrt{2}}(|00\rangle+|11\rangle$) \cite{Werner1989} as the atoms initial state. Werner state is a X-shape entangled state when $p> \frac{1}{3}$. Here we set $p=\frac{2}{3}$.

When $\nu=1$, the metric (\ref{csm}) describes a flat spacetime absent of the cosmic string. According to the following properties of Bessel function \cite{Andrews1999}
 \begin{align}
 \sum_{n=-\infty}^{\infty}J^{2}_{|n|}(x)=1,\quad \sum_{n=-\infty}^{\infty}J_{|n|+1}(x)J_{|n|-1}(x)=0,
\end{align}
we has
\begin{align}
&f^{(11)}_{11}=f^{(11)}_{22}=f^{(11)}_{33}=1,\nonumber\\
&g^{(12)}_{11}=g^{(12)}_{22}=\frac{3 [(L^2 \omega ^2-1) \sin (L \omega )+L \omega  \cos (L \omega )]}{2 L^3 \omega ^3},\nonumber\\
&g^{(12)}_{33}=\frac{3 [\sin (L \omega )-L \omega  \cos (L \omega )]}{L^3 \omega ^3},\nonumber\\
&g^{(12)}_{13}=0. \label{mks}
\end{align}
From above equations, we know that the results of Minkowski spacetime are recovered \cite{Huang2019}.
 \begin{figure}
\centering
\includegraphics[height=4cm,width=6cm]{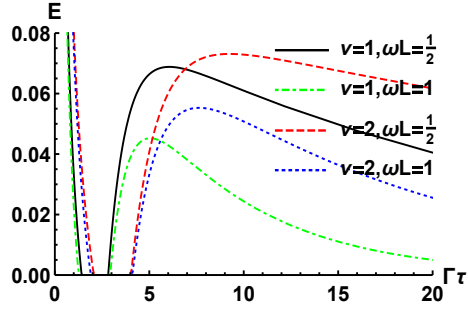}
\caption{\label{figure1} Entanglement as function of $\Gamma\tau $ with different atomic separations for $ \omega r=1/2$, and two atoms in cosmic string spacetime and Minkowski spacetime are polarized along isotropic direction ($\mathbf{\hat{d}}^{(1)}=\mathbf{\hat{d}}^{(2)}=(1/{\sqrt{3}}, 1/{\sqrt{3}}, 1/{\sqrt{3}})$).}
\end{figure}
From Fig. \ref{figure1}, it can be seen that in both spacetime, there exist time gap where entanglement vanishes and entanglement lifetime becomes longer when atomic separation become smaller. However, compared with the free Minkowski spacetime, entanglement in cosmic string spacetime decays more quickly in the beginning, and revives later. Entanglement lifetime  in cosmic string spacetime is longer than Minkowski spacetime when two-atom distances in both spacetime are the same.

 \begin{figure}
\centering
\subfigure[]{\includegraphics[height=4cm,width=6cm]{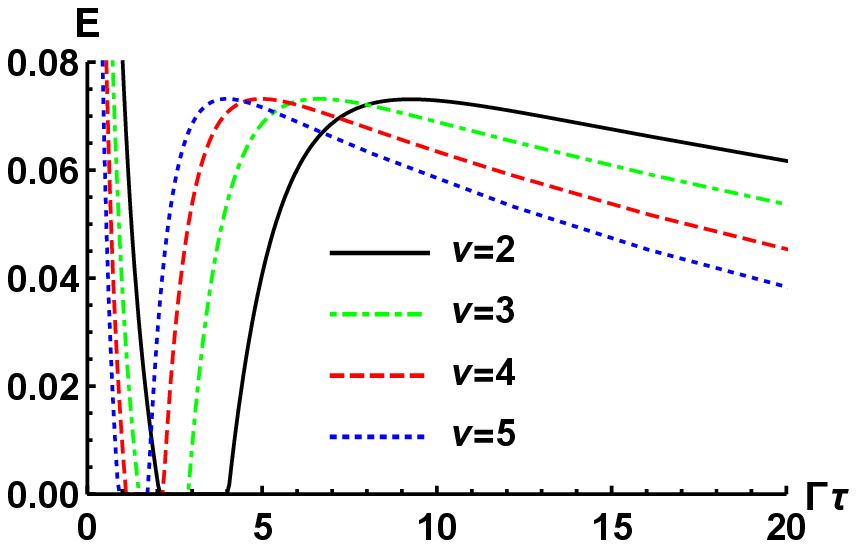}}
\subfigure[]{\includegraphics[height=4cm,width=6cm]{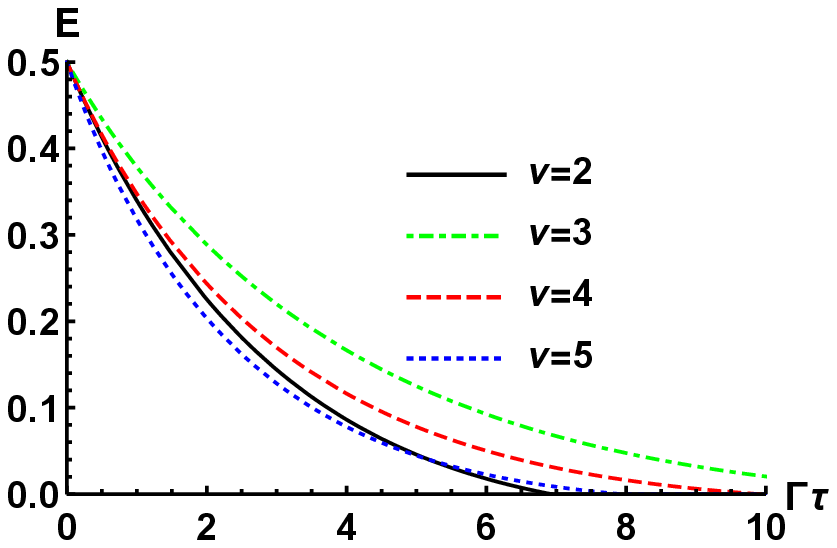}}
\caption{\label{figure2} Entanglement as function of $\Gamma\tau $ with different values of deficit angle parameter for $ \omega L=1/2$ and $ \omega r=1/2$, (a) two atoms are polarized along isotropic direction ($\mathbf{\hat{d}}^{(1)}=\mathbf{\hat{d}}^{(2)}=(1/{\sqrt{3}}, 1/{\sqrt{3}}, 1/{\sqrt{3}})$), (b) an atom is polarized along radial direction ($\mathbf{\hat{d}}^{(1)}=(1,0,0)$) and the other atom is polarized along tangential direction ($\mathbf{\hat{d}}^{(2)}=(0,1,0)$).}
\end{figure}
From Fig. \ref{figure2} (a), it can be found that bigger value of deficit angle parameter, entanglement lifetime becomes shorter, and entanglement decreases more rapidly at first, but sudden birth happens earlier and the gap becomes narrower. From Fig. \ref{figure2} (b), we can see when an atom is polarized along the positive radial direction and the other atom along the positive tangential direction, entanglement has different behaviors.
Entanglement shows a monotone decrease with evolution time, and entanglement and its lifetime does not decreases monotonously with increasing value of deficit angle parameter.

When $\nu>1$ and $ r\rightarrow0$  (two atoms lay on the string), according to the property of Bessel function
\begin{equation}
J_{x}(0)=
\begin{cases}
0& x>0\\
1& x=0
\end{cases},
\end{equation}
one obtain
\begin{align}
&f^{(11)}_{11}=f^{(11)}_{22}=0,\quad f^{(11)}_{33}=\nu,\nonumber\\
&g^{(12)}_{11}=g^{(12)}_{22}=0,\quad g^{(12)}_{33}=\frac{3 \nu [\sin (L \omega )-L \omega  \cos (L \omega )]}{L^3 \omega ^3},\nonumber\\
&g^{(12)}_{13}=0.
\end{align}
According to above equations, we get
\begin{align}
&A=\frac{\Gamma\nu}{4}|\hat{d}^{(1)}_3|^{2},\nonumber\\
&B=\frac{\Gamma\nu}{4}|\hat{d}^{(2)}_3|^{2},\nonumber\\
&C=\frac{3\Gamma \nu [\sin (L \omega )-L \omega  \cos (L \omega )]}{4L^3 \omega ^3}\hat{d}^{(1)}_3 \hat{d}^{(2)}_3. \label{zas}
\end{align}
From above equations, we know that when there art not atoms polarized along axial direction, the system does not affected by the electromagnetic fluctuation and the system seems like a closed system.
 \begin{figure}
\centering
\subfigure[]{\includegraphics[height=4cm,width=6cm]{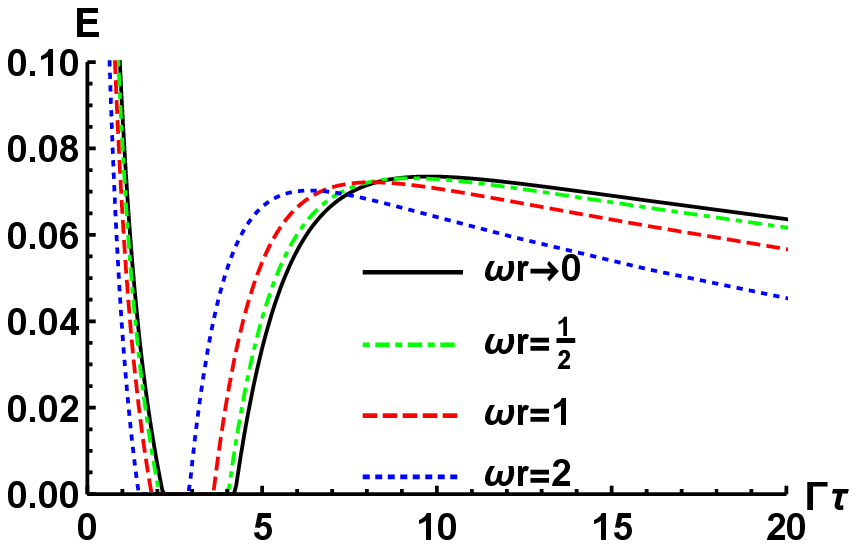}}
\subfigure[]{\includegraphics[height=4cm,width=6cm]{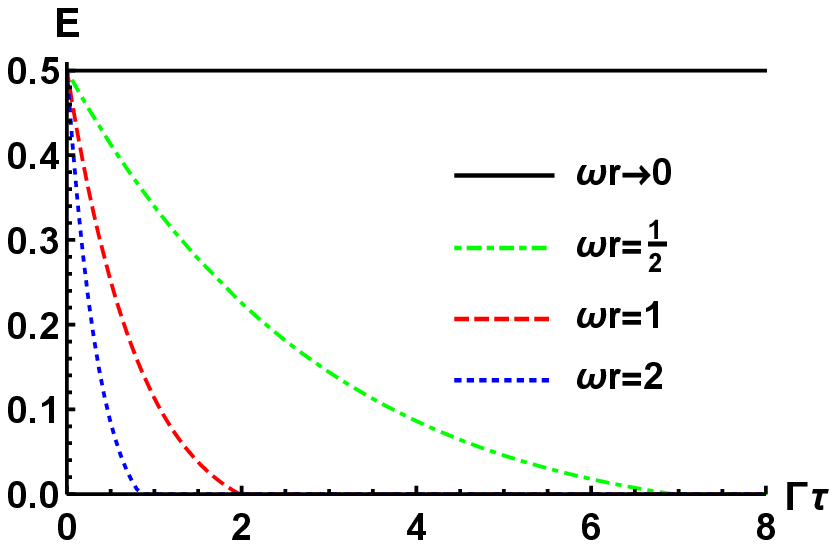}}
\caption{\label{figure3} Entanglement as function of $\Gamma\tau $ with different atom-string distances for $ \omega L=1/2$ and $ \nu =2$, (a) two atoms are polarized along isotropic direction ($\mathbf{\hat{d}}^{(1)}=\mathbf{\hat{d}}^{(2)}=(1/{\sqrt{3}}, 1/{\sqrt{3}}, 1/{\sqrt{3}})$), (b) an atom is polarized along radial direction ($\mathbf{\hat{d}}^{(1)}=(1,0,0)$) and the other atom is polarized along tangential direction ($\mathbf{\hat{d}}^{(2)}=(0,1,0)$).}
\end{figure}
 \begin{figure}
\centering
\subfigure[]{\includegraphics[height=4cm,width=6cm]{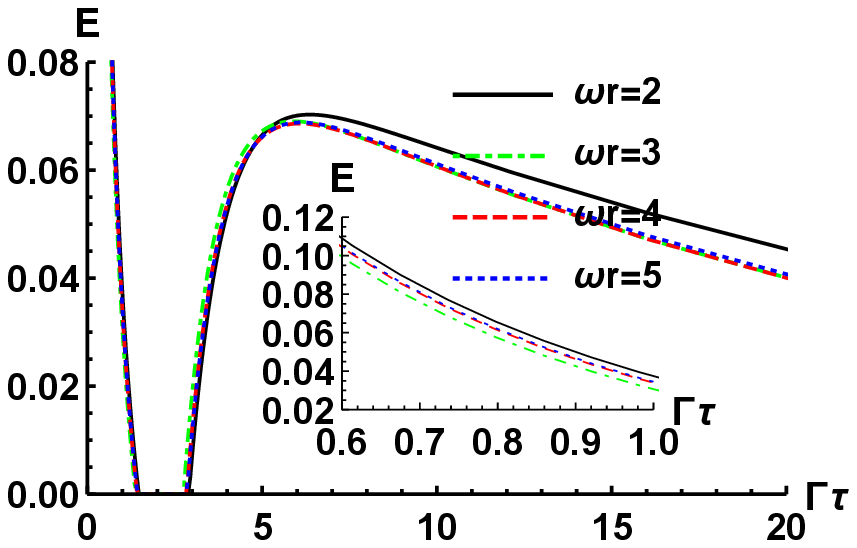}}
\subfigure[]{\includegraphics[height=4cm,width=6cm]{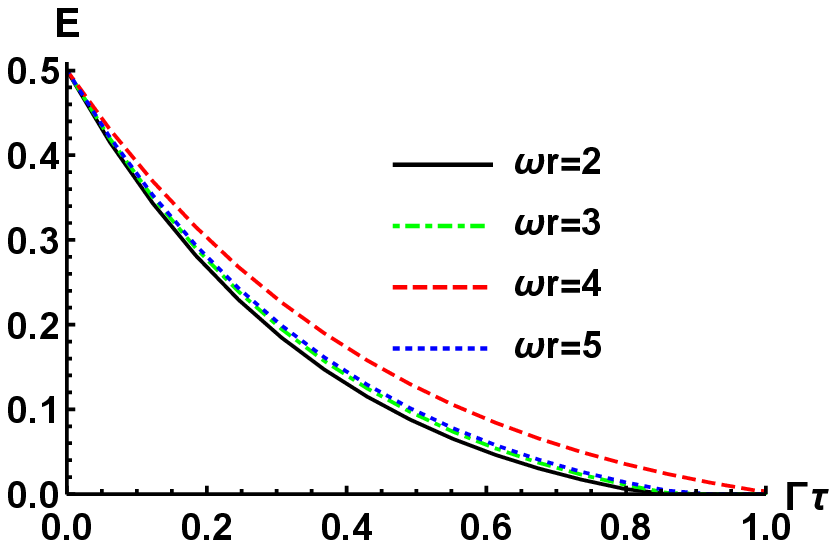}}
\caption{\label{figure4} Entanglement as function of $\Gamma\tau $ with different atom-string distances for $ \omega L=1/2$ and $ \nu =2$, (a) two atoms are polarized along isotropic direction ($\mathbf{\hat{d}}^{(1)}=\mathbf{\hat{d}}^{(2)}=(1/{\sqrt{3}}, 1/{\sqrt{3}}, 1/{\sqrt{3}})$), (b) an atom is polarized along radial direction ($\mathbf{\hat{d}}^{(1)}=(1,0,0)$) and the other atom is polarized along tangential direction ($\mathbf{\hat{d}}^{(2)}=(0,1,0)$).}
\end{figure}
Fig. \ref{figure3} shows that when two atoms relatively approach the string, entanglement decays more slowly and maintains more longer time; as the Eq. (\ref{zas}) reveals, entanglement is freezing when atoms are not polarized along axial direction (see Fig. \ref{figure3} (b)). When atom-string distance relatively large (see Fig.  \ref{figure4}), entanglement exhibits oscillations as atom-string distance varies. Especially when atoms are far away from the string ($r\rightarrow \infty$), Eq. (\ref{mks}) can be exactly obtained, which means the results in Minkowski are restored because the atoms are not affected by the cosmic string for this case. Entanglement has different behaviors for different polarization directions.

 \begin{figure}
\centering
\includegraphics[height=4cm,width=6cm]{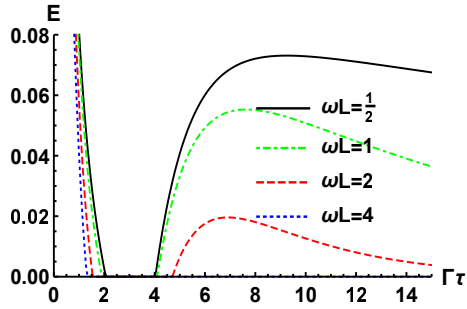}
\caption{\label{figure5} Entanglement as function of $\Gamma\tau $ with different two-atom distances for $ \omega r=1/2$ and $ \nu =2$, two atoms are polarized along isotropic direction ($\mathbf{\hat{d}}^{(1)}=\mathbf{\hat{d}}^{(2)}=(1/{\sqrt{3}}, 1/{\sqrt{3}}, 1/{\sqrt{3}})$).}
\end{figure}
From Fig. \ref{figure5}, it can be observed that when two-atom distance becomes smaller, entanglement declines more slowly in the beginning, the dark period of entanglement becomes shorter and entanglement can persist in longer time; particularly, when two-atom distance is relatively big, entanglement revival no longer occurs. In this sense, we can say small atomic separation can protect entanglement. When an atom is polarized along radial direction and the other atom is polarized along tangential direction, from Eq. (\ref{fts1}), we know that entanglement evolution has nothing to do with interatomic distance.

\section{Conclusion}\label{S4}
In this work, we have analyzed the entanglement behaviors for two atoms weakly coupled with a bath of fluctuating electromagnetic field in the cosmic string spacetime. We find that entanglement behaviors depend on polarization direction of the atoms, interatomic separation, atom-string distance and value of deficit angle parameter. It is shown that the asymptotic equilibrium state is a pure separable state, which indicates entanglement can not persevere in for a long evolution time. When deficit angle parameter $\nu=1$ and the atoms are far away from the string, the results reduce to that of Minkowski spacetime. For different values of deficit angle parameter, entanglement has different behaviors depending on various parameters. When atoms get close to the string, entanglement can be efficiently protected; in particular, when atoms lie on the string and are not polarized along axial
direction, entanglement is freezed, which is similar to the boundary effect \cite{Huang2019}. When atom-sting distance is relatively large, entanglement presents oscillating behaviors as atom-string distance changes. When two-atom separation is small, entanglement has better performance. Various parameters give us greater freedom to steer entanglement behaviors, and this would be profitable for us detecting the structure and property of the cosmic string spacetime and distinguishing different spacetime.

\end{document}